# Metallic rugate structures for perfect absorbers in visible and near-infrared regions

Shiwei Shu [1, 2] and Yang Yang Li [1, 2, *]

[1] Department of Physics and Materials Science, [2] Centre for Functional Photonics, City University of Hong Kong, 83 Tat Chee Av., Kowloon, Hong Kong

* E-mail: yangli@cityu.edu.hk, Phone: +852 3442 7810, Fax: +852 3442 0538

## ABSTRACT

Metallic rugate structures are theoretically investigated for achieving perfect absorption in the visible and near-infrared regions. Our model builds on nanoporous metal films whose porosity (volume fraction of voids) follows a sine-wave along the film thickness. By setting the initial phase of porosity at the top surface as 0, perfect absorption is obtained. The impacts of various structural parameters on the characteristic absorption behaviors are studied. Furthermore, multiple peaks or bands with high-absorption can be achieved by integrating several periodicities in one structure. The rugate absorbers show perfect or near-perfect absorption for TE and TM polarizations and large incident angles.

Rugate structure[1] is a type of optical film whose refractive index varies continuously (mostly sinusoidally) along the film thickness direction. In classical optics, rugate structures are generally used as “mirrors” for their resonant reflection peaks which are not affected by the initial phase of their periodic structures. Dielectric rugate structures have been intensively investigated, whereas few studies have focused on metal-based rugate structures. The little attention on metal-based rugate structures is possibly due to the fact that metal has considerable absorption in the visible and near-infrared regions, giving the illusion that the characteristic reflection responses of rugate structures will be weaken by using metal. Here, in this study, we report the surprising discoveries that the optical responses of metal rugate structures show strong dependency on their structural phase at the top surface and perfect absorption can be achieved.

Lately, high efficiency absorbers have been receiving lots of attention for their great promise in a wide range of applications such as solar cells, photodetectors, sensors, nanoimaging devices, and thermal emitters.[2-5] Various designs have been proposed and studied,[2-7] which mainly focus on patterning metal surfaces with periodic structural features. The difficulties in designing high efficiency absorbers lie in how to minimize the polarization-dependency and maximize the working range of incident angles, particularly for the visible and infrared light. Moreover, current designs generally expensive and complex nanofabrication techniques (e.g., E-beam lithography) which are limited to small sample sizes and flat substrates.

Here we report the theoretical study of metal rugate structures as perfect absorbers in the visible and near infrared regions. Moreover, the novel rugate absorber design eliminates any application of insulating dielectric spacer layers which are often present in the current absorber designs[2-4,6,7] and thus fully inherits many desirable properties of metals, e.g., high electrical and thermal conductivities, making them attractive for multi-functional applications. Regarding the manufacturing method, the novel rugate absorbers can be potentially constructed by the reported convenient electrochemical method[8] that is highly compatible with various substrate geometries and large sample sizes.

The designed metal-based rugate absorber is a nanoporous film with the porosity (volume fraction of pores), $p$, varying sinusoidally as a function of the film depth $z$:

$$p = p_a + A\sin(2\pi z / d + \varphi) \qquad (1)$$

The initial phase of the porosity sine wave at the film top surface is denoted as $\varphi$. The periodicity, amplitude of the porosity, average porosity, and the period number are denoted as $d$, $A$, $p_a$, and $m$, respectively. As will be shown later, perfect absorbers can be obtained when $\varphi$=0. Therefore, rugate absorbers are designed following eq. 2 (Fig. 1):

$$p = p_a + A\sin(2\pi z / d) \qquad (2)$$

The pores are designed to be significantly smaller than the wavelength of the incident light so that the effective medium theory can be readily applied for the simulation purposes. Other than that, there are no specific requirements on the size, shape, or distribution patterns of the pores, e.g., the pores can be randomly distributed. To simplify the calculation, the influence of the underlying dielectric substrate is neglected by setting the refractive index of the substrate to be unity.

Two methods were used to calculate the absorption spectra of the proposed metal-based rugate structures. The first method combined the Maxwell-Garnett effective medium theory (MGEMT) and the scattering matrix method (SMM), and the second is the FDTD method. We used the Lorentz-Drude (LD) model to describe metals in both methods.[9,10]

For the first method, one period of the rugate structures was divided into $j$ sub-layers with equal thickness, the average porosity of the $i^{th}$ sub-layer, $p_i$, was described as:

$$p_i = \frac{\int_{z_{i-1}}^{z_i} \left( p_a + A\sin(2\pi z / d) \right) dz}{d/m}, \ (i = 1, 2, ..., j) \tag{3}$$

where $z_i = id / j$ . Considering the feature size is much smaller than the wavelength concerned, each sub-layer was treated as a homogeneous material with effective permittivity $\varepsilon_{eff}$ according to the MGEMT:

$$\frac{\varepsilon_{eff} - \varepsilon_{air}}{\varepsilon_{eff} + 2\varepsilon_{air}} = (1 - p_i) \frac{\varepsilon_{metal} - \varepsilon_{air}}{\varepsilon_{metal} + 2\varepsilon_{air}} \tag{4}$$

Meanwhile, the FDTD method (commercial software EastFDTD) was used to check the accuracy of the results obtained from SMM. For the FDTD method, a square lattice of metal nanorods which stand vertically on a dielectric substrate was used to model the metallic rugate structures (Fig. S1). The periodicity in x-y plane was set to be 50 nm. The porosity of the film varies sinusoidally along the film thickness direction (the z axis) following eq. 1. The refractive index of the underlying dielectric substrate was set to be one.

Simulation study reveals that the optical responses of metal (e.g., Ni, W, or Ti) rugate structures are highly sensitive to $\varphi$, the initial phase of porosity at the film top surface (Figs. 2 and S2). Remarkable high absorption can be achieved with $\varphi = 0^o$, whereas an absorption dip, or a reflection peak, can be achieved with $\varphi = 180^o$. Furthermore, when $\varphi = 90^o$ or $\varphi = 270^o$, a pair of absorption peak and dip coexist closely within a sub-wavelength range. This is for the first time that dependency of the optical responses on the initial structural phase is observed on rugate structures. For comparison, the rugate structures reported to date are generally made from dielectrics and always display resonance reflection peaks regardless of the structural phase. The discovered strong phase dependency suggests the Fano-resonance in the metal rugate structures, which will be investigated in more depth in a later report. In this report, we focus on applying metal rugate structures as high efficiency absorbers.

The simulation results of the SMM and the FDTD method show good agreement with each other, e.g., both methods showed perfect or near-perfect absorption for Ni- or W-based rugate absorbers (Fig. S3). For conciseness, mainly the simulation results from the SMM method are presented.

We further compared the absorption behaviors of flat non-porous metal films, porous metal films with unvaried porosity along the film thickness direction, and rugate absorbers, all of the same film thickness (Figs. S4). Clearly, porosity in metals is helpful to greatly improve metal's absorption, whereas the specific periodic porosity profiles of rugate absorbers enables perfect absorption at characteristic wavelengths.

Further simulation work shows that the characteristic absorption of the rugate absorber can be adjusted by tuning its structural features, such as, the periodicity $d$, the average porosity $p_a$, the porosity amplitude $A$, and the period number $m$ (Figs. 3 and S5).

Apparently, the position of the characteristic absorption peak can be red-shifted by increasing *d* or decreasing $p_a$ (Figs. 3a-b and S5a-b). Interestingly, the porosity amplitude *A* appears to greatly affect the shape of the characteristic absorption peaks (Figs. 3c and S5c). By adjusting *A*, the peak contour can be transformed from a symmetric one to a highly asymmetric one that resembles the classic asymmetric Fano resonance line-shape. Regarding the impact of the period number *m* (Figs. 3d and S5d) the characteristic absorption peak appears only when $m \geq 2$, and reaches maximum height with *m* further increased to 8. This suggests that the high absorption efficiency of rugate absorbers is tightly related to their structural periodicity.

Furthermore, when increasing the refractive index of the dielectrics filled in the pores of the rugate absorber, the characteristic absorption peak red-shifts (Figs. 4e and S5e), not only suggesting another strategy to tune the optical response of the rugate absorber but demonstrating its potential application as optical sensors.

The rugate absorbers are insensitive to TE or TM polarization and possess good tolerance to the wide incident angle (Figs. 5 and S6). Taking a Ni-based rugate absorber for example, the maximum absorption drops only slightly from unity to 0.95 with the incident angle increased from $0^o$ to $50^o$ for TE polarization, and remains perfect from $0^o$ to $40^o$ for TM polarization. Interestingly, for TM polarization, when the incident angle is increased from $0^o$ to $60^o$, the absorption baselines go up over the entire studied spectrums for both Ni and W rugate absorbers, whereas, more apparently for the W rugate absorber, the maximum absorption also increases.

Another advantage of the rugate absorbers lies in the fact that multiple absorption peaks[6] or widened absorption bands can be achieved by integrating several single-peak rugate absorbers into one structure (Figs. 5 and S7). For example, a "dual" rugate absorber can be simply constructed by integrating two single rugate absorbers of the same $p_a$ and $A$ but different periodicity (namely $d_1$ and $d_2$). The porosity along the film thickness direction, $p_{dual}$, of the "dual" rugate absorber can be set as:

$$p_{dual} = 0.5*\left(p_a + A\sin(2\pi z / d_1)\right) + 0.5*\left(p_a + A\sin(2\pi z / d_2)\right) \tag{5}$$

Following this strategy, dual peaks or widened bands of high absorption which either overlap with or span the characteristic absorption peaks of the two component rugate absorbers can be readily achieved (Figs. 5 and S7).

Metal rugate structures as a novel type of high efficiency absorbers have been investigated. The porosity of the proposed rugate absorber follows a sine wave along the film thickness direction with the initial phase set as 0 at the top surface. Perfect absorption can be achieved in the visible and near-infrared regions for not only both TE and TM polarizations but also large incident angles. Furthermore, the absorption peaks can be fine-tuned by adjusting the various structural parameters of the metallic rugate structures and multiple absorption peaks or absorption bands can be achieved by integrating several periodicities in one rugate structure. It should be pointed out that according to the Kirchhoff's law[11] which states that at thermal equilibrium, the emissivity of a body equals to its absorptivity at every wavelength, the proposed absorbers can also be used as high-efficiency emitters. Moreover, the proposed design of metal-based rugate absorbers can fully utilize the many desirable attributes of metallic materials, which may potentially lead to a wide range of novel multi-functional optical devices.

**ACKNOWLEDGEMENT** This work is supported by City University of Hong Kong (Projects 7008080 and 9667030).

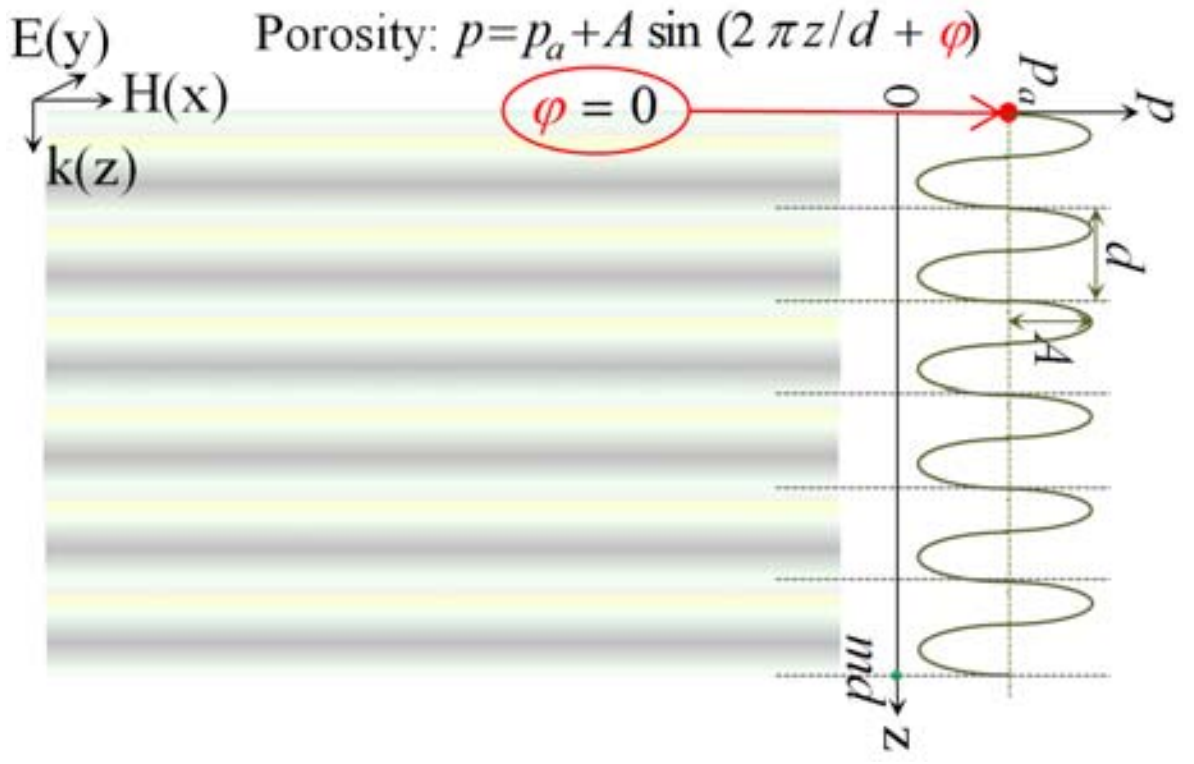

Fig. 1 Structural model of metal-based rugate structures used in the SMM method.

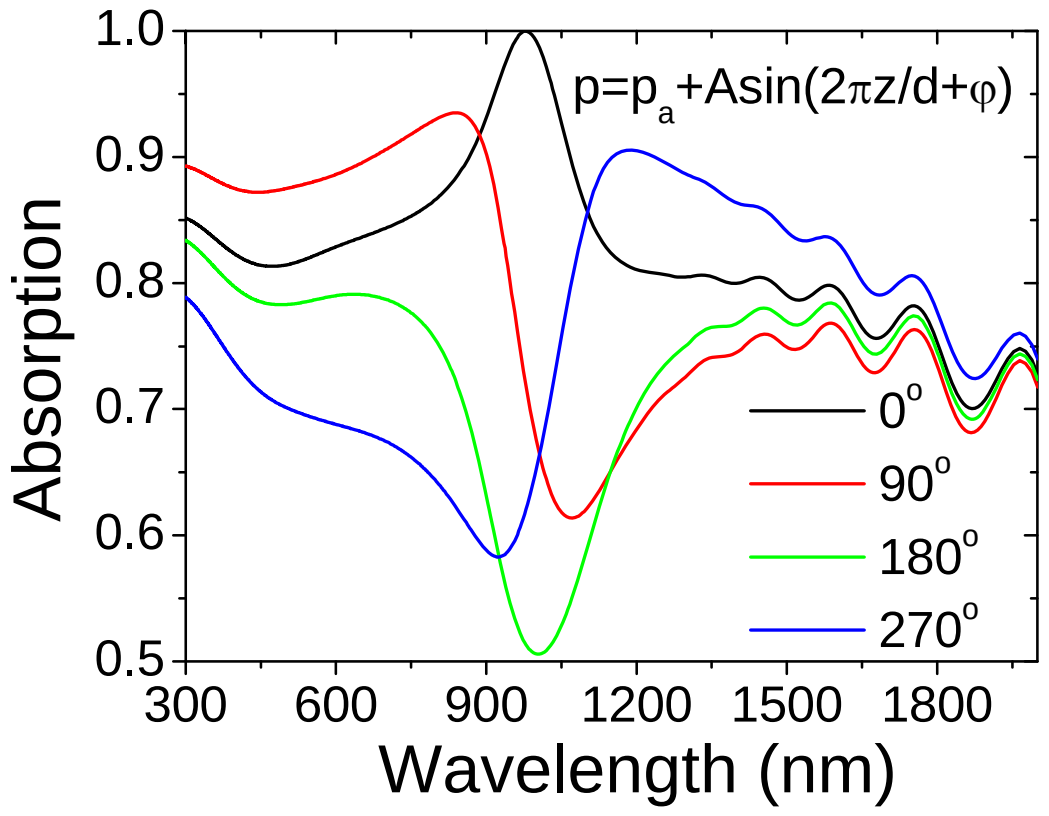

Fig. 2 Normal incident TE polarization absorption spectra of Ni-based rugate structures with different initial phase of porosity at the top film surface, $\varphi$ ($p_a$ = 40%, $A$ = 10%, $d$ = 200 nm, and $m$ = 16).

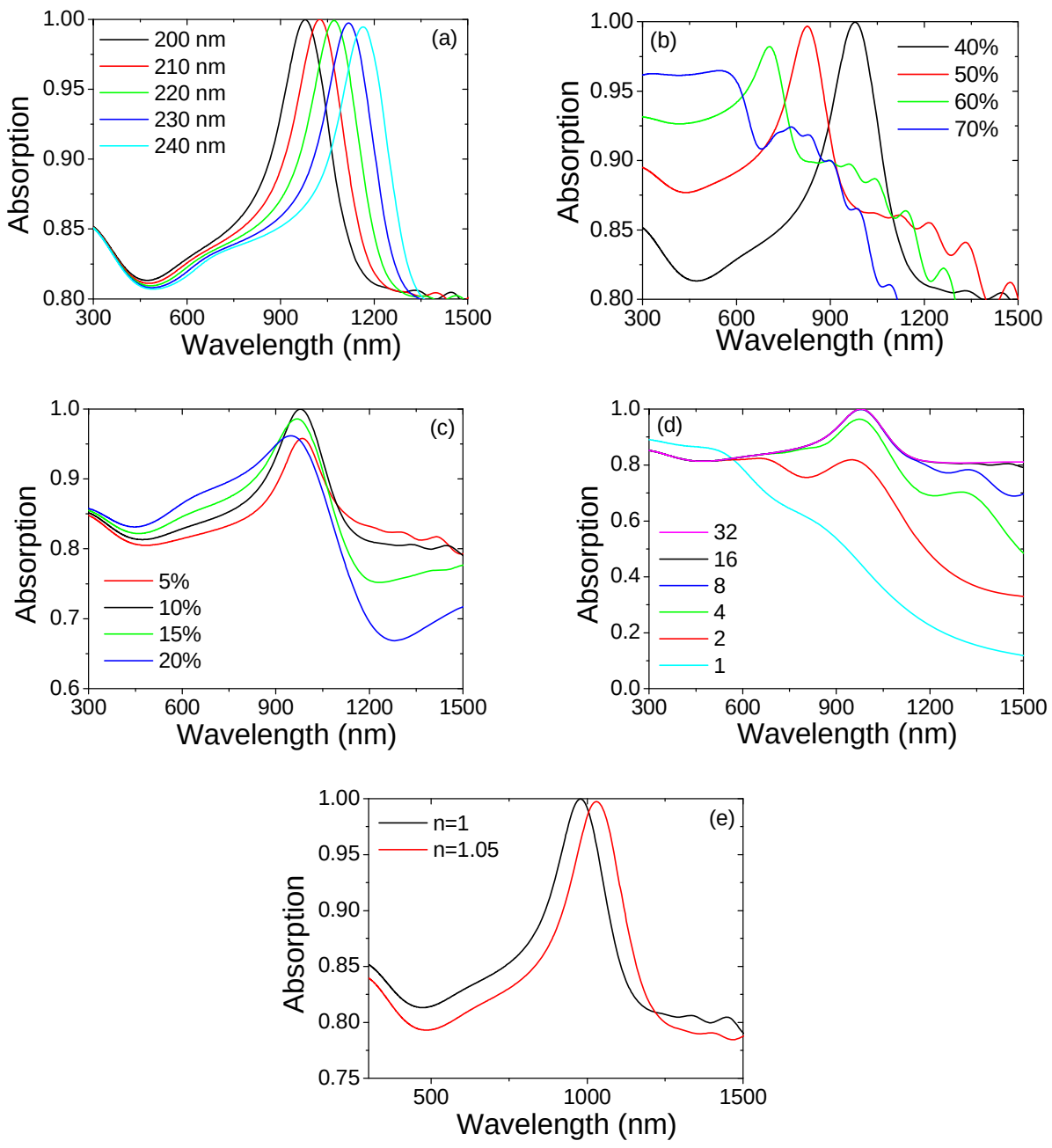

Fig. 3 Normal incident TE polarization absorption spectra of a Ni rugate absorber ($d$ =200 nm, $p_a$ = 40%, $A$ = 10%, and $m$ = 16). The impacts of the following parameters are studied: a) $d$, ranging from 200 nm to 240 nm; b) $p_a$, ranging from 40% to 70%; c) $A$, ranging from 5% to 20%; d) $m$, ranging from 1 to 32; e) the refractive index $n$ of the material filling the pores, changed from 1 to 1.1.

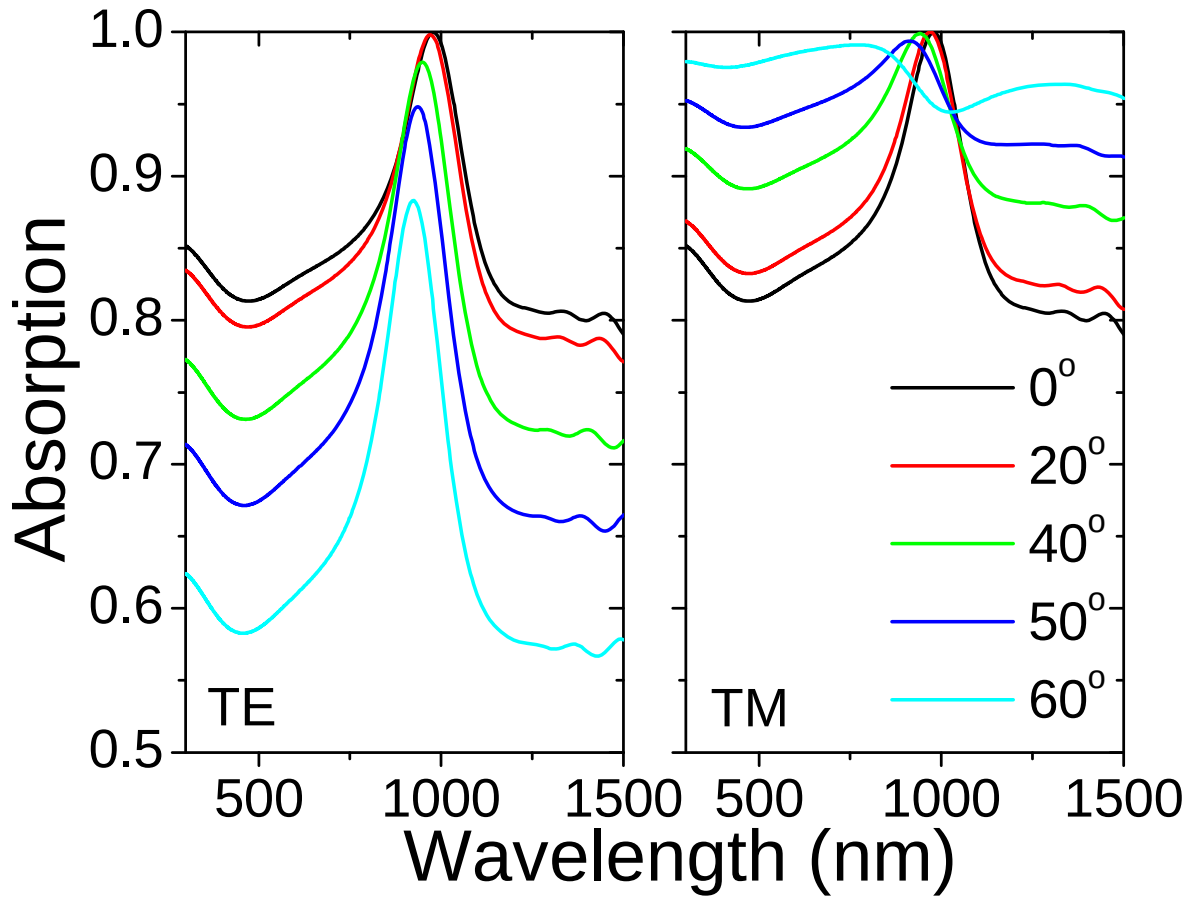

Fig. 4 (a) Absorption spectra of a Ni rugate absorber ($d$ = 200 nm, $p_a$ = 40%, $A$ = 10%, and $m$ = 16) with TE and TM polarizations and different incident angles.

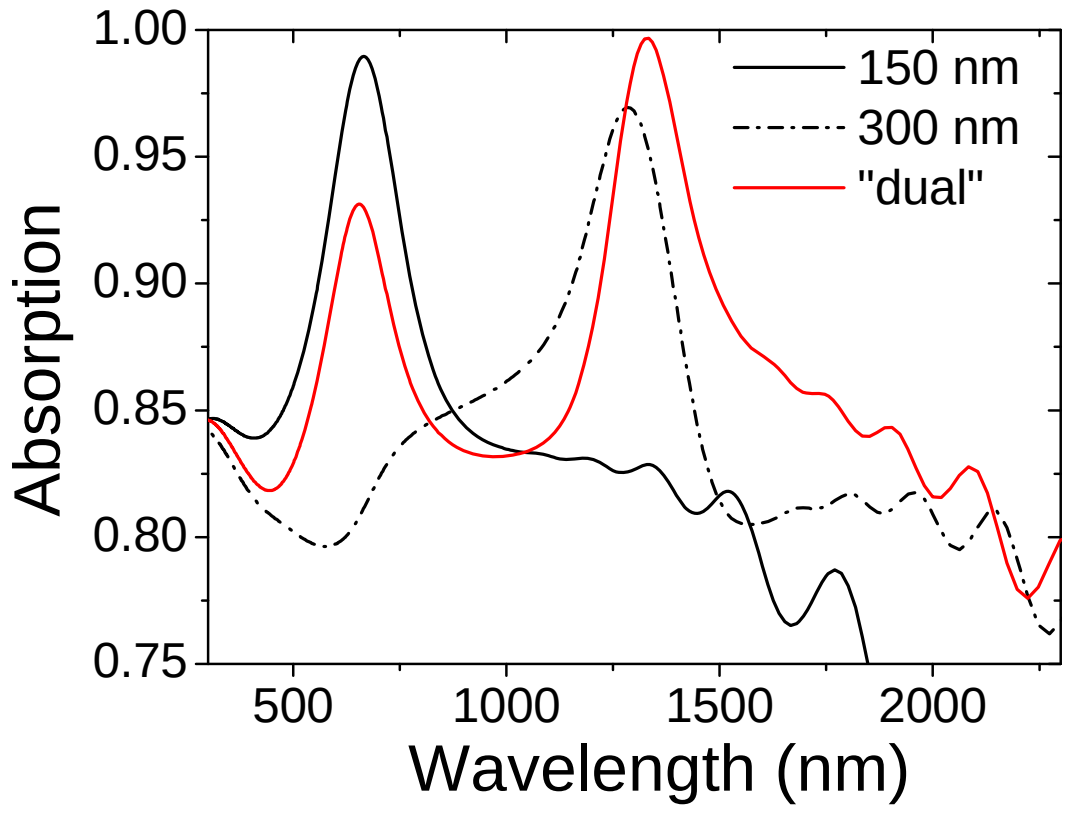

Fig. 5 Normal incident TE polarization absorption spectra of "dual" rugate absorbers constructed following eq. 5 and their component single rugate absorbers calculated using the FDTD method: Ni-based, $d_1$ = 150 nm, $d_2$ =300 nm, $d_{dual}$ = 300 nm, $p_a$ = 40%, $A$ = 10%, and $m$ = 16.